\documentclass[aps,prd,letterpaper,preprintnumbers,floatfix,superscriptaddress,twocolumn]{revtex4}
\pdfoutput=1 
\usepackage{graphicx,dcolumn,bm,epsfig,cancel,hyperref}
\usepackage{multirow}
\usepackage{amsmath}
\usepackage{amssymb, times}
\usepackage[utf8]{inputenc}
\usepackage[normalem]{ulem}
\usepackage{color}

\newcommand{\comment}[1]{{}}

\newcommand{\bbaa}[0]{b\bar{b}\gamma\gamma}

\newcommand{\bbet}{b\bar{b}+\not\!\! E_T}
\newcommand{\aaet}{\gamma\gamma+\not\!\! E_T}
\newcommand{\hhbbet}{hh\to b\bar{b}+\not\!\! E_T}
\newcommand{\hhaaet}{hh\to \gamma\gamma+\not\!\! E_T}
\newcommand{\met}{\not\!\! E_T}

\begin{document}
\title{
Dark and Bright Signatures of Di-Higgs Production
}
\author{Alexandre Alves}
\email{aalves@unifesp.br}
\affiliation{Departamento de Física, Universidade Federal de São Paulo, UNIFESP, Diadema, Brazil}

\author{Tathagata Ghosh}
\email{tghosh@hawaii.edu}
\affiliation{Department of Physics \& Astronomy, University of Hawaii, Honolulu, HI 96822, USA}

\author{Farinaldo S. Queiroz}
\email{farinaldo.queiroz@iip.ufrn.br}
\affiliation{International Institute of Physics, Universidade Federal do Rio Grande do Norte,
Campus Universitário, Lagoa Nova, Natal-RN 59078-970, Brazil}

\begin{abstract}
  If the Higgs boson decays to a pair of invisible particles, the number of di-Higgs events, where each Higgs decay into Standard Model (SM) particles, are reduced by a factor of two-third taking into account the current LHC bound on invisible decay width of the Higgs boson. We investigate the sensitivity of the upcoming high luminosity run of the LHC to di-Higgs production and subsequent decay to dark matter in the context of the singlet scalar extension of the SM augmented by a fermionic dark matter in the dark and bright channel $\aaet$. Once systematic uncertainties on background yields are considered, this dark and bright channel presents competitive limits than $\bbet$ after a careful tuning of the kinematical cuts that raise the signal over background ratio. We further show that in a multivariate analysis, for an invisible branching fraction as low as $\sim 10$\%, we obtain stronger bounds for the Higgs trilinear coupling from the $\aaet$ channel compared to the $\bbaa$ final state. 
 Finally, we demonstrate that the three channels $\aaet$, $\bbet$ and $\bbaa$, complement each other in the search for di-Higgs production with non-SM trilinear couplings when an invisible decay mode is present.
\end{abstract}


\maketitle

\section{ Introduction}
After the discovery of the Higgs boson at the LHC~\cite{lhc_h_ATLAS,lhc_h_CMS}, and the confirmation that this spin-0 particle plays a role in the electroweak symmetry breaking (EWSB) mechanism~\cite{Aad:2015mxa,Sirunyan:2017tqd,Sirunyan:2019twz}, 
the LHC still has a long road to unravel the details of the Higgs boson self-interactions by measuring its trilinear and quartic couplings. These measurements are key in the
understanding of the stability of the Higgs potential and the nature of the electroweak phase transition (EWPT).
The Standard Model (SM) Higgs potential, by itself, is metastable and cannot trigger a first order EWPT either. Additional scalar fields present in a beyond the SM model (BSM) can facilitate stabilizing the Higgs potential and also can trigger EWPT with profound implications for the matter-antimatter asymmetry. Thus, the stability of the vacuum and a 1$^{st}$ order EWPT make the Higgs self-interactions possible targets for signals of BSM physics. At the HL-LHC, such signals might reveal themselves either as deviations of the trilinear Higgs coupling as compared to its SM value or new resonances. 

Another physics goal of the upcoming runs of the LHC is to search for dark matter (DM) and its possible connection to the Higgs. The upcoming high luminosity run of the LHC (HL-LHC) has an enormous potential to discover or exclude DM models, including the Higgs portal models~\cite{Baek:2011aa,Baek:2012uj,Craig:2014lda,Baek:2015lna,Ko:2016ybp,Silveira:1985rk,McDonald:1993ex,Barger:2008jx,Cline:2013gha,Djouadi:2011aa,Hambye:2008bq,Baek:2012se,Baek:2013dwa,Kamon:2017yfx,Dutta:2017sod} where Higgs bosons couple to the DM field. Such interaction is expected if the DM acquires its mass via the EWSB mechanism. 

The LHC experiments have searched for invisibly decaying Higgs bosons and the current most stringent bound for its branching ratio into an invisible dark state is provided by the CMS collaboration, which is 19\%~\cite{Sirunyan:2018owy}. In contrast, in all the searches for di-Higgs by CMS and ATLAS, they do not consider any significant beyond the SM decay channel of the Higgs in addition to the SM modes, which is, of course,
a reasonable assumption given the state of affairs. However, if the Higgs boson indeed possesses an invisible decay channel with branching ratio, $\hbox{BR}_{inv}$, then decay rates of the Higgs to all the SM channels will be reduced by $(1-\hbox{BR}_{inv})$.  Hence, the di-Higgs production rate with both Higgs bosons decaying to the SM particles will go down by a factor of $(1-\hbox{BR}_{inv})^2$, which is around 0.65 taking into account the current LHC bound. Thus, all the discovery prospects of di-Higgs will also drop nearly by the same factor. In this scenario, looking for channels where one of the Higgs bosons decaying invisibly becomes an important task. 

Recently, the $\hhbbet$~\cite{Banerjee:2016nzb} channel was studied in the context of the SM and a singlet scalar extension of it, called xSM~\cite{Espinosa:2011ax, Profumo:2007wc, Profumo:2014opa}. xSM is an economical model with interesting phenomenological consequences like detectable gravitational waves signals from strongly 1$^{st}$ order EWPT, and a new massive scalar that decays to $hh$. Among the $hh$ channels involving $\met$, this is the final state with the largest expected number of events but also with the largest backgrounds, mainly from $b\bar{b}Z$ and $Zh$. In the unlikely case where systematic effects can be neglected, this channel might reach around $4\sigma$ of statistical significance in the SM case using a Boosted Decision Trees (BDT) algorithm. However, its signal-to-background ratio is tiny, and once even very small systematic backgrounds are taken into account, that significance drops below $1\sigma$.

In this paper, we study the mode $\hhaaet$, a dark and bright decay of the di-Higgs system, and explore non-SM trilinear couplings in the context of xSM~\cite{Espinosa:2011ax, Profumo:2007wc, Profumo:2014opa} augmented by a fermionic DM. If the DM is a scalar particle, the $Z_2$ symmetry required to stabilize the DM will preclude cubic terms in the scalar potential, which facilitate to trigger a 1$^{st}$ order EWPT.  However, in such models 1$^{st}$ order EWPT can still be induced via zero-temperature loop effects, thermally driven scenarios, and modifications to the scalar potential at tree level by renormalizable or non-renormalizable operators (see~\cite{Curtin:2014jma} and references therein). On the other hand, for a vector DM one needs an extra scalar to generate its mass via spontaneous symmetry breaking~\cite{Hambye:2008bq,Baek:2012se,Baek:2013dwa,Kamon:2017yfx,Dutta:2017sod}. A detailed study on these intricacies will be presented elsewhere. 

We focus on the part of the xSM parameters space that respects several experimental and theoretical constraints, and where new heavy Higgs bosons are not expected to be observable at the LHC but shift the SM Higgs trilinear coupling away from its SM value. In this case, our results can be useful to constrain other models with non-SM trilinear couplings. We do not restrict ourselves to points that might deliver a strong gravitational wave signal in future space-based interferometers; however, the parameters space of interest in the present study is expected to cover that where a strongly $1^{st}$ order EWPT might occur~\cite{Alves:2018jsw}.


We also compare this channel's performance with $\bbet$ and $\bbaa$ final states. Even though the expected number of signal events is much lower than $\bbet$, the backgrounds are also much smaller, and a more favorable $S/B$ is helpful to tame the systematic uncertainties for similar signal significances. We show that by carefully tuning the kinematic cuts using a Gaussian Process algorithm, the photons plus missing energy channel has a competitive reach both for discovery as well as exclusion compared to the $\bbet$ channel once realistic systematic uncertainties are included in the computation. Moreover, using BDT along with tuned cuts, we show that $\aaet$ becomes competitive with $\bbaa$ in terms of signal significance, with invisible branching ratios close to the current LHC bound for non-SM trilinear couplings which make the $hh$ cross sections large.

Besides, we want to emphasize that estimating the reach of $\aaet$ is also important in view of the combination with another channel like $\bbet$. Combining several search channels is often possible in experimental studies which take systematic effects and correlations into account. This is explicitly shown in the case of double Higgs production into SM channels in Ref.~\cite{Cepeda:2019klc}. 


Our paper is organized as follows. In Section \ref{sec:model}, we introduce our xSM model assumptions. We impose the constraints on the xSM parameter space for our collider analysis in Section \ref{sec:const}. In Section~\ref{sec:aaet}, we lay out our strategy to analyze the $\aaet$ final state. The results of the cut-and-count and multivariate analyses on the $\aaet$ channel are presented in Sections~\ref{sec:cuts} and \ref{sec:BDT}, respectively. Finally, we conclude in Section~\ref{sec:conclusions}.



\section{\label{sec:model} Model assumptions}

An invisibly decaying SM Higgs boson might arise in Higgs portal models where the Higgs couples either directly to scalar ($S$) or vector ($V$) dark matter fields through renormalizable interactions like $|S|^2|H|^2$ and $V_\mu V^\mu |H|^2$, respectively. In the case of fermionic DM, renormalizable interactions with the Higgs field arises only by the addition of an extra scalar. When such additional scalar acquires a vacuum expectation value (vev), the DM mass generation through spontaneous symmetry breaking occurs.

For example, consider the simple extension where one real gauge singlet is added to the Higgs potential, the so-called xSM~\cite{Espinosa:2011ax, Profumo:2007wc, Profumo:2014opa}. Now consider that this extra scalar mediates the interaction with DM as follows
\begin{eqnarray}
  {\cal L} &=& V(H,S) + y_{\chi}S\bar{\chi}\chi\nonumber \\
  V(H,S) &=& -\mu^2 H^{\dagger} H + \lambda (H^{\dagger}H)^2 
  + \frac{a_1}{2} H^{\dagger} H S \nonumber \\
  &+& \frac{a_2}{2} H^{\dagger} H S^2 + \frac{b_2}{2} S^2 + \frac{b_3}{3} S^{3} + \frac{b_4}{4}S^4
\end{eqnarray}
where $H^T=(0,h+v_{EW})/\sqrt{2}$ is the SM Higgs doublet and $S=s+v_S$ the new scalar, where $v_{EW}$ and $v_S$ are the vev of the SM Higgs and the new singlet scalar, respectively. The Higgs-DM interactions read, after EWSB,
\begin{equation}
  {\cal L}_{int} \supset -\sin\theta\frac{m_\chi}{v_S}h_1\bar{\chi}\chi+\cos\theta\frac{m_\chi}{v_S}h_2\bar{\chi}\chi,
\end{equation}
where the mixing angle between $s$ and $h$ is $\theta$, $h_1$ is the 125 GeV Higgs and $h_2$ the heavier one
\begin{equation}
h = \cos\theta\; h_1-\sin\theta\; h_2,\;\; s = \cos\theta\; h_2+\sin\theta\; h_1\;\; .
\end{equation} 
The fermionic dark matter is assumed to get its mass entirely from the new scalar vev; thus it's mass, and Yukawa coupling is related by $m_\chi = y_\chi v_S$. We highlight that throughout we will not impose the relic density constraints because for the couplings adopted here one may reproduce the right relic dark matter density either via the thermal freeze-out or non-thermal production of dark matter \cite{Arcadi:2017kky}. If a dominant non-thermal production is assumed the bounds stemming from direct and indirect detection might be circumvented \cite{Camargo:2019ukv}. 

Due to the mixing of scalars, the SM Higgs boson can decay to a DM pair but with a $\sin\theta$ suppression. Otherwise, even if no mixing of the scalars is allowed, the SM Higgs decays to DM at one-loop mediated by the heavy scalar. These decays are possible due to the scalars self-interactions
\begin{equation}
    {\cal L}_{int} \supset \sum_{\{ijk\}=1,2} \lambda_{ijk}h_ih_jh_k+\sum_{\{ijkl\}=1,2} \lambda_{ijkl}h_ih_jh_kh_l\; .
    \label{L:self}
\end{equation}
The heavy Higgs decay to dark matter, by its turn, can have a sizable decay branching ratio, suppressing its decay to SM Higgs pairs.

The mixing with the new scalar induces deviations in the SM Higgs trilinear coupling given by
\begin{equation}
\kappa_\lambda \equiv \frac{\lambda_{111}}{\lambda_{111}^{SM}}\approx 1+\theta^2\left(-\frac{3}{2}+\frac{2m^2_{h_2}-2b_3v_S-4b_4v_S^2}{m^2_{h_1}}\right), 
\label{eq:dev}
\end{equation}
where $b_3$ and $b_4$ are free parameters, and $\theta$ is assumed to be small in view of the current bound $ |\sin\theta| \lesssim 0.20$  coming from the 1-loop correction to the $W$ boson mass~\cite{Robens:2015gla}.

\begin{figure}[t]
  \includegraphics[width=0.95\columnwidth]{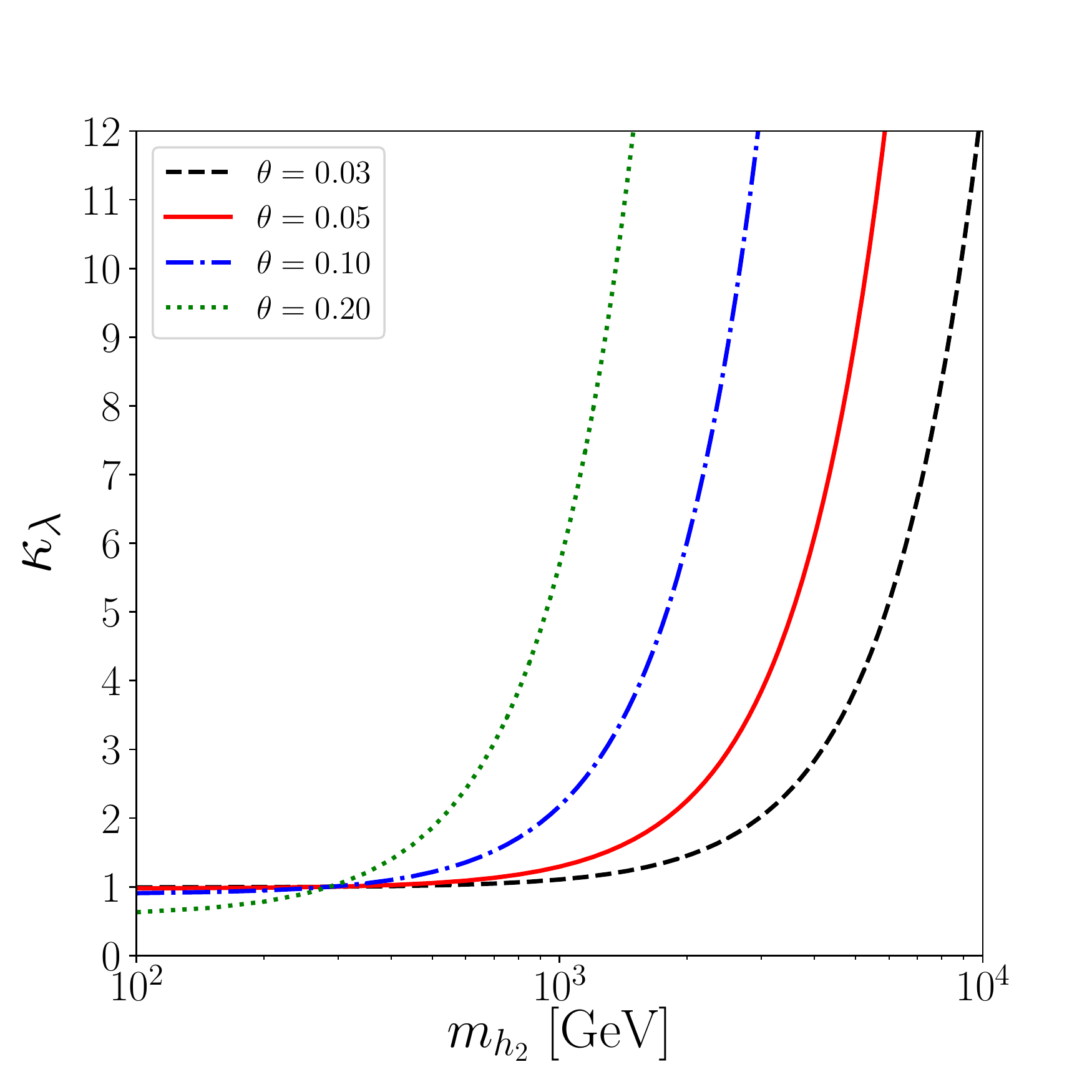} 
\caption{\label{fig:lam}
Deviation of the SM Higgs trilinear coupling as a function of the $h_2$ mass in four different Higgs mixing angle scenarios. We fixed $v_S=0.1v_{EW}$, $b_3=10v_{EW}$ and $b_4=5$ in this plot. The shift in the trilinear coupling is not sensitive to these parameters for large $h_2$ masses in the typical cases we are going to study.
}
\end{figure}

From Eq.(\ref{eq:dev}) above, we see that a heavy new scalar shifts $\lambda_{111}$ away from its SM value toward large positive values. In Fig.~(\ref{fig:lam}) we display $\kappa_\lambda$ as function of the $h_2$ mass for some fixed mixing angles and other parameters. We see that scenarios with large shifts for $m_{h_2}>500$ GeV are possible. For small mixings, however, large $\kappa_\lambda$ will occur for $h_2$ masses in the region of TeVs. For $h_2$ masses smaller than $\sim 1$ TeV, $\kappa_\lambda$ is more sensitive to variations in $b_3$, $b_4$ and $v_S$, but for the typical parameters we are interested in, the picture does not change much. 

A dark and bright signature, $pp\to \gamma\gamma+\met$,  might receive several contributions in xSM. When $\theta$ is non-vanishing, tree-level $h_1\to\chi\bar{\chi}$ and 1-loop $h_2\to\gamma\gamma$ are possible. Besides the double Higgs production $pp\to h_1h_1\to\gamma\gamma+\chi\bar{\chi}$, more contributions to the process $\gamma\gamma+\met$ arise: (1) $pp\to h_2\to h_1h_1$, (2) $pp\to h_1h_2$, (3) $pp\to h_2h_2$, all of them involving trilinear couplings, and (4) $pp\to h_ih_jh_k$, $pp\to h_ih_j$ and $pp\to \chi\bar{\chi}+h_{1,2}$ involving quartic couplings. The $h_1h_1$ production with an enhanced trilinear $\lambda_{111}$ coupling, when $\theta$ is small and $h_2$ is heavy, is the dominant contribution while the other contributions are expected to be negligible. We explain them in details below.  

%

\begin{itemize}
\item[(1)] \underline{$pp\to h_2\to h_1h_1$}. This is the resonant contribution where the heavy Higgs boson $h_2$ decays to a pair of SM Higgs bosons. This contribution might produce hard photons and large missing transverse momentum but it is suppressed by the scalar mixing angle as $\sin^2\theta$ in the gluon fusion production of $h_2$. The decay into dark matter, by its turn, diminishes the branching to $h_1h_1$. The cross section is low for large $h_2$ masses.

\item[(2)] \underline{$pp\to h_1h_2$}. There are three contributions here, actually, with either $h_1$ or $h_2$ off-mass shell, and a box diagram. The $pp\to h_1^*\to h_1h_2$ is expected to have a too low cross section. The other channel, $pp\to h_2^*\to h_1h_2$, is suppressed by $\sin^2\theta$. Note that, in these cases, DM can be produced directly in the decay of $h_2$. The branching ratio of $h_2\to\gamma\gamma$ is suppressed by the mixing angle just like the effective $h_2 gg$ coupling. The box diagram is also suppressed by $\sin\theta$.

\item[(3)] \underline{$pp\to h_2h_2$}. This is analogous to $pp\to h_1h_1$ with a triangle and a box contribution where $h_2$ couples to the top quark. The triangle contribution is suppressed by $\sin\theta$ whereas the box contribution by $\sin^2\theta$. Moreover, the photons should be produced by the $h_2$ decay. There is another contribution where $h_1^*\to h_2h_2$, but again, its cross section is expected to be very small.

\item[(4)] Triple Higgs production, for example, $pp \to h_2/h^*_2 \to h_1h_1h_1$, also produces photons plus missing energy but with a tiny production cross-section and an additional suppression coming from the invisible branching ratio of the third Higgs. Double production involving quartic couplings occurs at the 2-loop level only. Finally, $pp\to \chi\bar{\chi}+h_1$, with the quartic interaction $\lambda_{2111}$ is also possible but again at the 2-loop level. 
All these contributions are expected to be small too.

\end{itemize}

If the mixing angle is not too small and $h_2$ is not too heavy, it is expected that the resonant contribution (1) substantially enhances the $h_1h_1$ production cross section and helps to disentangle the background events efficiently due to the production of hard photons and missing energy. The other channels might also contribute to this scenario. However, if the mixing angle is too small or $m_{h_2}$ is of $\mathcal{O}$(TeV), or both, even the resonant contribution will be negligible. In this case, the sole contribution is $pp\to h_1h_1\to\gamma\gamma+\chi\bar{\chi}$ and, as we discussed before, a significant deviation in the trilinear coupling might be possible. Otherwise, model parameters that lead to large Higgs masses and small shifts will be tough to probe at the HL-LHC. We will adopt a conservative and less model-dependent approach and suppose that contributions from $h_2$ are negligible. 

Our results can thus apply to models where the trilinear coupling is shifted away from its SM value.
There exist other models, like the Composite Higgs boson model~\cite{Contino:2010rs, Giudice:2007fh, Contino:2010mh}, and the Two Higgs Doublet Models \cite{Arhrib:2009hc}, which also predict non-standard trilinear couplings which might enhance double Higgs production. If the Higgs boson decays to dark matter, our results might apply.

In this paper, we are going to explore the double SM-like Higgs $h_1$ production and decay to an invisible channel plus two photons where the only possibility for an enhanced cross section is by changing the triple Higgs coupling $\lambda_{111}$. As is well known, the triangle and the box contribution to $gg\to h_1h_1$ interfere destructively, but for $|\lambda_{111}|$ sufficiently large, the production cross section grows beyond the SM value. Models with new heavy states contributing sizeably to $h_1h_1$ production should be easier to explore once they are more discernible from backgrounds. For this moment, we stick to the hardest scenario with no resonances or new contributions to the double Higgs production. 


The motivation to explore a final state with large missing energy is that once the Higgs boson possesses an invisible decay mode with branching ratio $\hbox{BR}_{inv}$, branching ratios of all other decay channels drop by a factor $(1-\hbox{BR}_{inv})$. In particular, for double Higgs production, any final state containing just the SM particles will be suppressed by a factor $(1-\hbox{BR}_{inv})^2$. The current LHC bound on the Higgs decay to invisible particles (19\%~\cite{Sirunyan:2018owy}) means that any SM only final state studied so far, will have 1/3 less number of events. Consequently, the LHC reach for di-Higgs production into SM channels would suffer a substantial depletion. Exploring channels with one of the Higgs boson decaying invisibly is an interesting exercise to evaluate their potential in the search for double Higgs production under the hypothesis of the existence of an invisible channel. In the case where the HL-LHC has no sensitivity to the dark matter decays, it is still essential to determine the potential of the collider to explore the classic di-Higgs channels, like $\bbaa$, with reduced branching ratios in scenarios with large trilinear couplings.

\section{\label{sec:const} Model Constraints}

 A number of theoretical and phenomenological constraints apply to this model. First, the scalar potential need to be bounded from below, which leads to the following conditions
\begin{equation}
\lambda > 0,\;\; b_4 > 0,\;\; a_2 \geq -2\sqrt{\lambda b_4}\; .
\label{eq:below}
\end{equation} 

Next, we impose that the potential be stable (at $T=0$) by solving the equations
\begin{eqnarray}
&& \frac{\partial V}{\partial h}=\frac{\partial V}{\partial s}=0\; , \\
&& \frac{\partial^2 V}{\partial h^2}>0,\frac{\partial^2 V}{\partial s^2}>0\; ,\\
&& \frac{\partial^2 V}{\partial h^2}\frac{\partial^2 V}{\partial s^2}-\left(\frac{\partial^2 V}{\partial h\partial s}\right)^2>0\; .
\label{eq:stable}
\end{eqnarray}

We also require that all the $2\to 2$ scattering amplitudes involving $W,Z,h_1$ and $h_2$ bosons respect perturbative unitarity at the high energy limit. Details of these calculations can be found in Ref.~\cite{Alves:2018jsw}.

From the experimental side, we take the constraints for the Higgs mixing angle~\cite{Carena:2018vpt}, the Higgs branching into invisible~\cite{Sirunyan:2018owy}, the current bound in the trilinear coupling shifts $\kappa_\lambda$~\cite{Cepeda:2019klc}, and $W$ mass corrections\cite{Lopez-Val:2014jva, Robens:2015gla}
\begin{eqnarray}
|\sin\theta| < 0.33,\;\; BR(h_1\to\hbox{invisible})<0.19,\nonumber\\ -4.7<\kappa_\lambda<12,\;\; \Delta M_W\in [-5\;\hbox{MeV},55\;\hbox{MeV}]\; .
\label{eq:vinc_exp}
\end{eqnarray}
For $|\sin\theta|<0.33$, $h_2$ masses up to a few TeV also respect constraints from oblique parameters $S,T$ and $U$~\cite{Baek:2011aa}. The $W$ mass bounds imply $|\sin\theta|<0.2$, effectively.

The branching fraction of the Higgs boson into fermionic DM (invisible) is given, in this model, by
\begin{eqnarray}
BR(h_1\to \chi\bar{\chi}) &=& \frac{\Gamma_{\chi\bar{\chi}}}{(1-\sin^2\theta)\Gamma_{SM}+\Gamma_{\chi\bar{\chi}}} \\
\Gamma_{\chi\bar{\chi}} & = & \frac{\sin^2\theta\; m_\chi^2}{8\pi v_S^2}\;m_{h_1}\left(1-\frac{4m_\chi^2}{m_{h_1}^2}\right)^{3/2}, 
\end{eqnarray}
where $\Gamma_{SM}=4.07$ MeV. The branching ratio of $h_1$ to any other SM particles is multiplied by $1-BR(h_1\to \chi\bar{\chi})$.


In our model, the DM-nucleus scattering is spin-independent and nearly isospin conserving. Thus, one can easily compute the dark matter-nucleon scattering cross section and compare with most stringent limits up-to-date that stem from the XENON1T experiment. In cm$^2$ we find the DM-nucleon spin-independent scattering cross section to be~\cite{Baek:2011aa,LopezHonorez:2012kv,Arcadi:2019lka},
\begin{equation}
\sigma_{SI} = 5\times 10^{-35}\times \left(\frac{y_\chi \sin(2\theta)\mu_{\chi N}}{v_S}\right)^2
\times\left(\frac{1}{m_{h_1}^2}-\frac{1}{m_{h_2}^2}\right)^2\;\; ,
\label{eq:xsecSI}
\end{equation}
where $\mu_{\chi N}$ is the reduced mass of the proton-DM system. We impose the latest XENON1T limits~\cite{Aprile:2018dbl} on the model. We point out that there are other competitive bounds in the literature stemming from other experiments, but we will adopt the XENON1T for being the most restrictive \cite{Akerib:2016vxi,Cui:2017nnn}.

It is known that fermionic DM Higgs portal models typically lead to an over-abundance of DM in the universe unless the DM mass lies very close to the threshold of resonant annihilations via one of the Higgs bosons~\cite{Baek:2012uj}. In our case, the SM Higgs and the new Higgs can play the role of mediators between the DM and the SM particles. We assume that the new Higgs is heavy. Thus, DM annihilations through $h_2$ are not efficient, unless we are near its resonance, but this will not happen here because the dark matter mass is sufficiently small. Because we want to explore the LHC prospects for Higgs into invisible states, we restrict the dark matter mass to 
\begin{equation}
m_{\chi} < 60\; \hbox{GeV}
\label{eq:xmass}
\end{equation}
and keep $y_\chi$ in the perturbative regime. As aforementioned we will assume that the correct relic dark matter density is somehow achieved invoking non-thermal effects that could bring down the abundance to its correct value as measured by PLANCK experiment \cite{Almeida:2018oid,Bernal:2018hjm,Borah:2019bdi}.

\begin{figure}[t]
  \includegraphics[width=1\columnwidth]{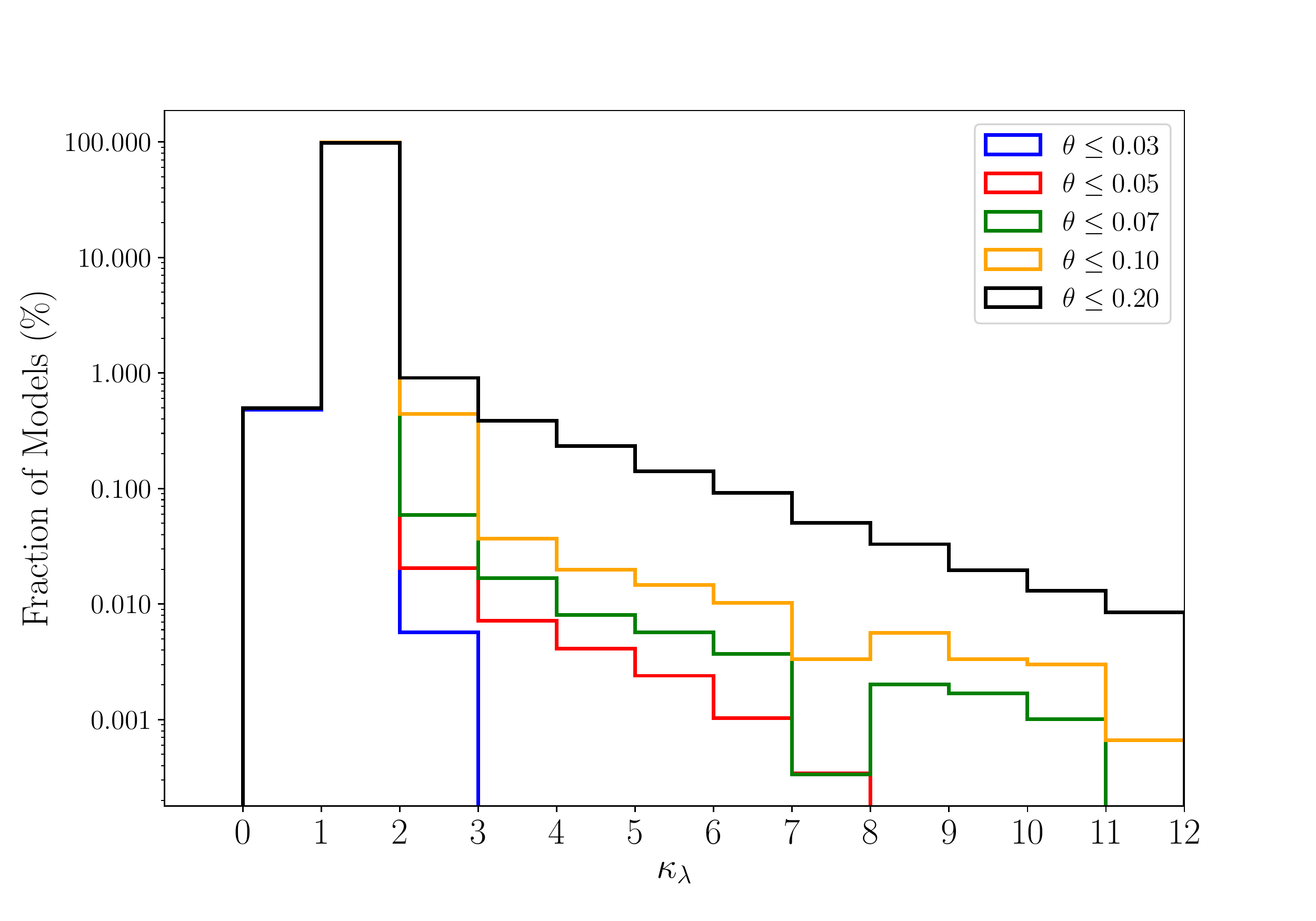} 
\caption{\label{fig:scan}
The fraction of models which satisfy the theoretical and experimental constraints of Eqs.~(\ref{eq:below}--\ref{eq:xmass}) is shown as a function of $\kappa_\lambda$, which is the shift in the SM Higgs trilinear coupling for five fixed mixing angles.
}
\end{figure}

With the constraints of Eqs.~(\ref{eq:below}--\ref{eq:xmass}) at hand, we scan over the relevant portion of the parameters $(\theta,b_3,b_4,v_S,m_{h_2},m_\chi)$ of the xSM augmented with the fermionic DM interactions. In Fig.~(\ref{fig:scan}), we show the number of models as function of $\kappa_\lambda$ which respect all the constraints in six scenarios from $\theta\leq 0.03$ to $\theta\leq 0.20$. As $\theta$ gets large, the number of models with large trilinear shifts increases, however, the majority of the models lie in the $1<\kappa_\lambda < 2$ region. 

\section{\label{sec:aaet} The search channel $\gamma\gamma+\met$}

The di-Higgs channel involving $\met$, which produces the most significant number of events is, naturally, $\bbet$ that was studied in Ref.~\cite{Banerjee:2016nzb}. In that work, a cut-and-count and a machine learning (ML) analysis were performed for both non-resonant and resonant scenarios within the xSM. The non-resonant study was performed for the SM trilinear coupling only, and $\sim 2.3\sigma$ and $\sim 4.3\sigma$ signal significances were estimated from the cut-and-count and the ML analysis, respectively. As we mentioned before, the study does not take systematic uncertainties into account and assume that errors are statistically dominated.
The caveat of those results is that the signal-to-background ratio, $S/B$, is tiny, amounting to just $0.03$ in the most powerful computation using BDT. The backgrounds to this channel include $b\bar{b}Z$, $b\bar{b}W$, $Zh_1$ and $t\bar{t}$. Once even a minimal uncertainty is taken into account in the number of background events, the significance drops below $1\sigma$. The situation in the resonant case is most promising being less sensitive to systematic effects.

The signal cross section of $\aaet$ suffers from the low branching ratio to photons but, the backgrounds to this channel are expected to be much lower compared to $\bbet$. The dominant ones are: (1) continuum $Z\gamma\gamma$, (2) $q\bar{q}\to Zh_1$, and (3) $gg\to Zh_1$ at 1-loop level, where the $Z$ boson decays to neutrinos. The reducible contributions $W\gamma\gamma$ and $Wh_1$ can be neglected after vetoing a hard charged lepton and imposing the full selection cuts in all the subsequent analysis.
In fact, we are going to show that despite starting with a much smaller signal cross section due to low $h \rightarrow \gamma \gamma$ branching ratio, we can achieve a signal-to-background ratio that is an order of magnitude higher than that of $\bbet$.
%
%

We simulate signal and background events at the 14 TeV LHC using \texttt{MadGraph5\_aMC@NLO\_v2.3.6}~\cite{Alwall:2014hca} at leading order. Higher order QCD corrections are included in total cross sections via appropriate K-factors given in Table~(\ref{tab:xsec}). The dark matter mass is fixed at 50 GeV in the simulations. The distributions depend weakly on the DM mass once the missing transverse momentum corresponds to the Higgs $p_T$ up to detector effects.
We use \texttt{Pythia8}~\cite{Sjostrand:2014zea} for hadronization and showering, and the detector simulation is performed by \texttt{Delphes3}~\cite{deFavereau:2013fsa}.
The number of SM signal and backgrounds events with QCD corrections, assuming 3 ab$^{-1}$, are shown in Table~(\ref{tab:xsec}) with the following basic selection criteria
\begin{equation}
p_T(\gamma) > 20\hbox{ GeV},\; |\eta(\gamma)|<2.5,\; \met > 20\hbox{ GeV},\;\Delta R_{\gamma\gamma}>0.5\; ,
\label{eq:basic}
\end{equation}
where the minimum transverse momentum within the fiducial volume of the electromagnetic calorimeters were required for the two photon hardest photons of the event. Here, $p_T(\gamma),\; \eta(\gamma),\;\Delta R_{\gamma\gamma}$ are the transverse momentum, rapidity and the distance between the photons in the $\eta\times \phi$ plane, respectively, while $\met$ is the missing transverse momentum of the event.
%
\begin{table}[t]
\centering
\begin{tabular}{c|c|c|c}
\hline\hline
 Signal($\lambda_{SM}$) & $Z\gamma\gamma$ & $qq\to Zh_1$ & $gg\to Zh_1$\\ 
\hline
37.2 & 3762 & 358.2 & 50.6 \\
\hline\\
2.27~\cite{deFlorian:2013uza} & 1.2~\cite{Alwall:2014hca} & 1.5~\cite{Brein:2003wg, Cepeda:2019klc} & 1.35~\cite{Cepeda:2019klc} \\
\hline\hline
\end{tabular}
\caption{In the first row, we display the number of events for 3 ab$^{-1}$ expected for the $\aaet$ signal, assuming SM trilinear coupling, and the backgrounds, after the basic selection cuts discussed in the Eq.~(\ref{eq:basic}). The second line shows the QCD K-factors used to normalize the signal and the background cross sections with respective references. 
}

\label{tab:xsec}
\end{table}
\section{\label{sec:cuts} Cut analysis and comparison with $\bbet$ and $\bbaa$ channels}

First of all, we want to show that once systematic uncertainties are taken into account, $\aaet$ presents better prospects than $\bbet$ at the HL-LHC. We are going to compare the signal significance of both channels for the same level of systematic effect despite this is not likely to be realistic. 
Jet energy calibration, b-tagging efficiency and a higher number of backgrounds with QCD radiation are some features which are expected to impact more severely the $\bbet$ channel~\cite{Sirunyan:2019mbp} while the much more precise photon identification of the detectors and non-QCD backgrounds favors $\aaet$. 

In order to maximize the signal significance in the photons plus missing energy mode, we tuned the kinematic cuts using \texttt{CutOptmize}~\cite{CutOptimize}, a \texttt{Python} package aimed to maximize the statistical significance in particle physics searches at colliders in cut-and-count and ML analysis using a Gaussian Process algorithm as implemented in \texttt{Hyperopt}~\cite{hyperopt}. Especially, \texttt{CutOptmize} learns to tame the systematic uncertainties in the number of background events by raising $S/B$. We did not attempt to optimize the bottoms channel but we believe that optimizing cuts can also raise the $S/B$ ratio making it less sensitive to a common systematic uncertainty, $\varepsilon_{sys}$. The number of signal and background events for $\bbet$ was taken from Ref.~\cite{Banerjee:2016nzb} and the significance metrics as well, which is given by
\begin{equation}
N_\sigma = \frac{S}{\sqrt{B+\varepsilon_{sys}^2(B^2+S^2)}}\;\; .
\label{eq:sig}
\end{equation}

\begin{figure}[t]
  \includegraphics[width=1\columnwidth]{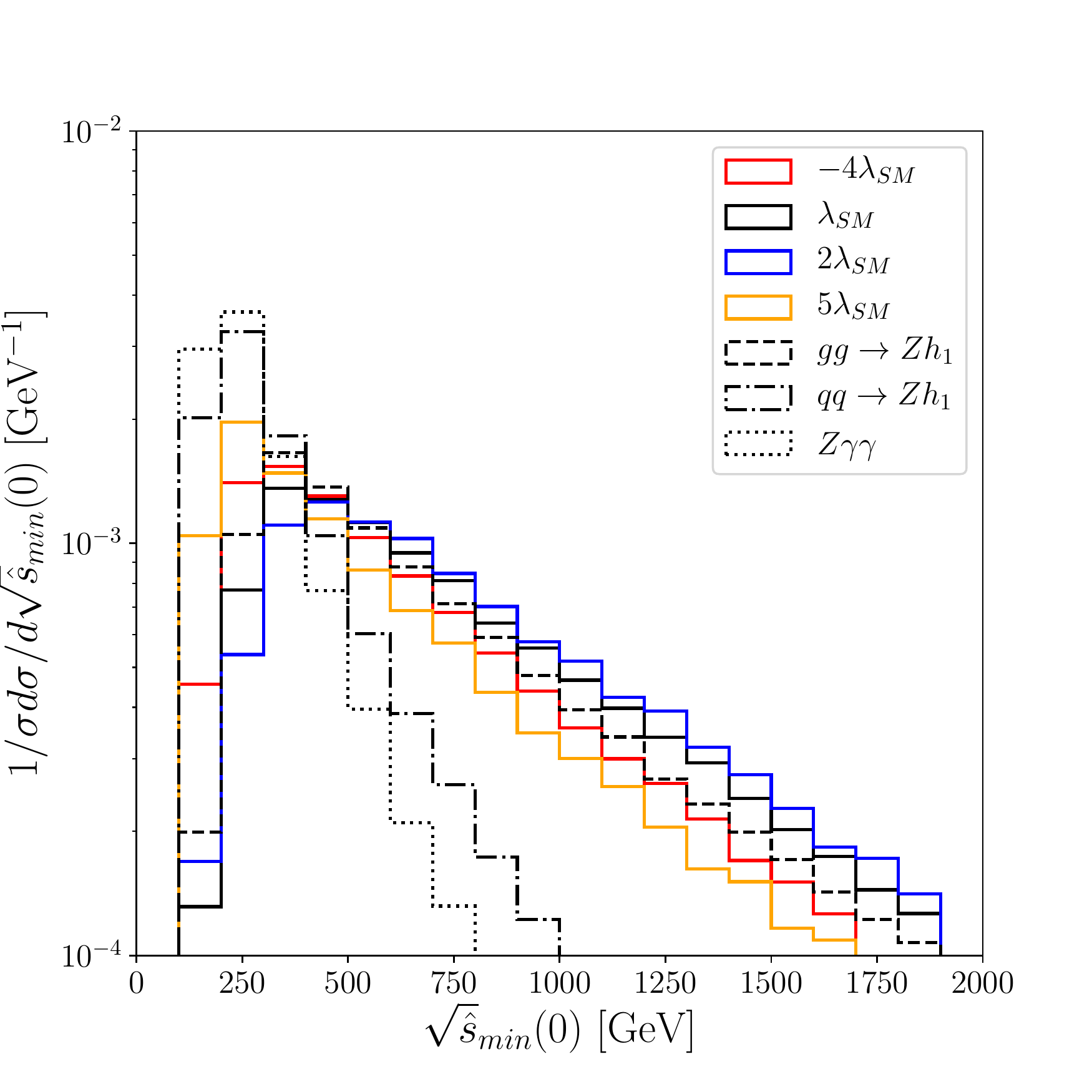}\\ 
  \includegraphics[width=1\columnwidth]{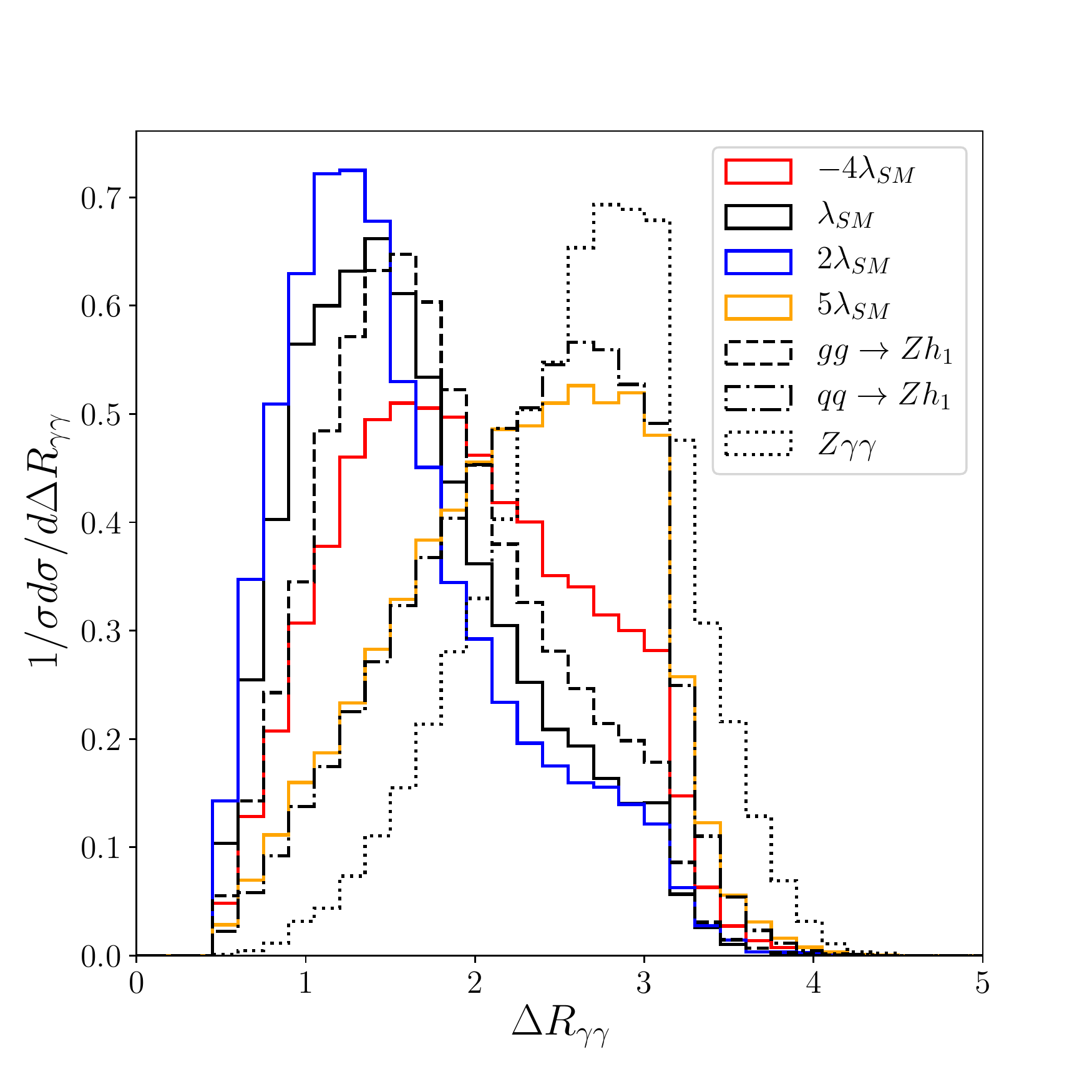}
\caption{\label{fig:kin}
The $\sqrt{\hat{s}}_{min}(0)$ distribution  (this the 11$^{th}$ variable defined below) is displayed at the upper panel, and the photons distance in $\eta\times\phi$ plane, $\Delta R_{\gamma\gamma}$, at the lower one.
}
\end{figure}
The kinematic variables used for cuts and the BDT classification are the following: 
\begin{itemize}
\item[(1)] the transverse momentum of the two leading photons: $p_{T_{\gamma_1}},\; p_{T_{\gamma_2}}$,
\item[(2)] the transverse momentum of the photons pair: $p_{T_{\gamma\gamma}}$,
\item[(3)] the missing transverse momentum: $\met$, 
\item[(4)] the mass of the photons pair: $M_{\gamma\gamma}$,
\item[(5)] the difference between the azimuthal angles of photons: $\Delta\phi_{\gamma\gamma}$, 
\item[(6)] the distance between the photons in the $\eta\times\phi$ plane: $\Delta R_{\gamma\gamma}$, 
\item[(7)] the angle between photons and missing transverse momentum vectors: $\Delta\phi(\vec{p_{T_{\gamma_1}}},\vec{\not\!\! p_T}),\;\Delta\phi(\vec{p_{T_{\gamma_2}}},\vec{\not\!\! p_T}),\;\Delta\phi(\vec{p_{T_{\gamma\gamma}}},\vec{\not\!\! p_T})$,
\item[(8)] the Barr variable~\cite{Barr:2005dz}, defined by: $\cos\theta^*=\tanh(\Delta\eta_{\gamma\gamma}/2)$,
\item[(9)] $MT_A=\sqrt{2p_{T_{\gamma\gamma}}\met-2\vec{p_{T_{\gamma\gamma}}}\cdot \vec{\not\!\! p_T}}$,
\item[(10)] $MT_B=\sqrt{2E_{\gamma\gamma}\met-2\vec{p_{T_{\gamma\gamma}}}\cdot \vec{\not\!\! p_T}}$ 
\item[(11)] $\sqrt{\hat{s}}_{min}(0)=\sqrt{E_{\gamma\gamma}^2-p_{L_{\gamma\gamma}}^2}+\met$~\cite{Konar:2008ei},
\item[(12)] the number of jets in the event: $N_j$ with $p_T>20$ GeV and $|\eta|<2.5$.
\end{itemize}

In Fig.~(\ref{fig:kin}), we show the signal and background distributions of the variables $\met$ and the distance between the photons, $\Delta R_{\gamma\gamma}$ at the upper and lower panel, respectively. The $\sqrt{\hat{s}}_{min}(0)$ spectrum displays a behavior which is observed in other kinematic variables with energy dimensions as well. They are: (1) the Higgsstrahlung $q\bar{q}\to Zh_1$ and the continuum $Z\gamma\gamma$ backgrounds are softer than the signals for trilinear couplings around the SM value, (2) the harder spectra occur for $\lambda_{111}$ closer to $\lambda_{SM}$, the values with the strongest destructive interference and, consequently, with smaller cross sections, (3) as $|\lambda_{111}|$ increases, the signal distributions get softer, (4) the $gg\to Zh_1$ background resembles the SM signal distributions more closely. The photons distance in $\eta\times\phi$ plane, $\Delta R_{\gamma\gamma}$, presents a similar behavior: the Higgsstrahlung $q\bar{q}\to Zh_1$ and the continuum $Z\gamma\gamma$ backgrounds are more easily distinguishable from the signals with $\lambda_{111}\sim \lambda_{SM}$, however, $gg\to Zh_1$ is, again, similar to the SM.

A note of caution is necessary at this point. For photons produced centrally in the detector, $|\eta_\gamma|<1.5$, and with energies up to $\sim 200$ GeV, the \texttt{Delphes3} parametrization of the photon energy resolution leads to a full width at half of maximum (FWHM) of the photons pair mass of 5 GeV which is roughly twice as large as the CMS and ATLAS resolutions~\cite{Cepeda:2019klc}. This difference can be attributed in part to the simplified approach to the photons simulation from \texttt{Delphes3} which neglects $e^+e^-$ conversions in the electromagnetic calorimeter. It is not clear, however, if the discrepancy has other sources. Our typical cut requirements are hard enough to select events with high energy though. For hard photons, the energy resolution is not an issue and we can trust the distributions generated by \texttt{Delphes3}. Moreover, we do not select events with too narrow $\gamma\gamma$ widths, limiting the distance to the Higgs mass to 5 GeV at least, large enough to accommodate the FWHM from \texttt{Delphes3}.
\begin{figure}[t]
  \includegraphics[width=.88\columnwidth]{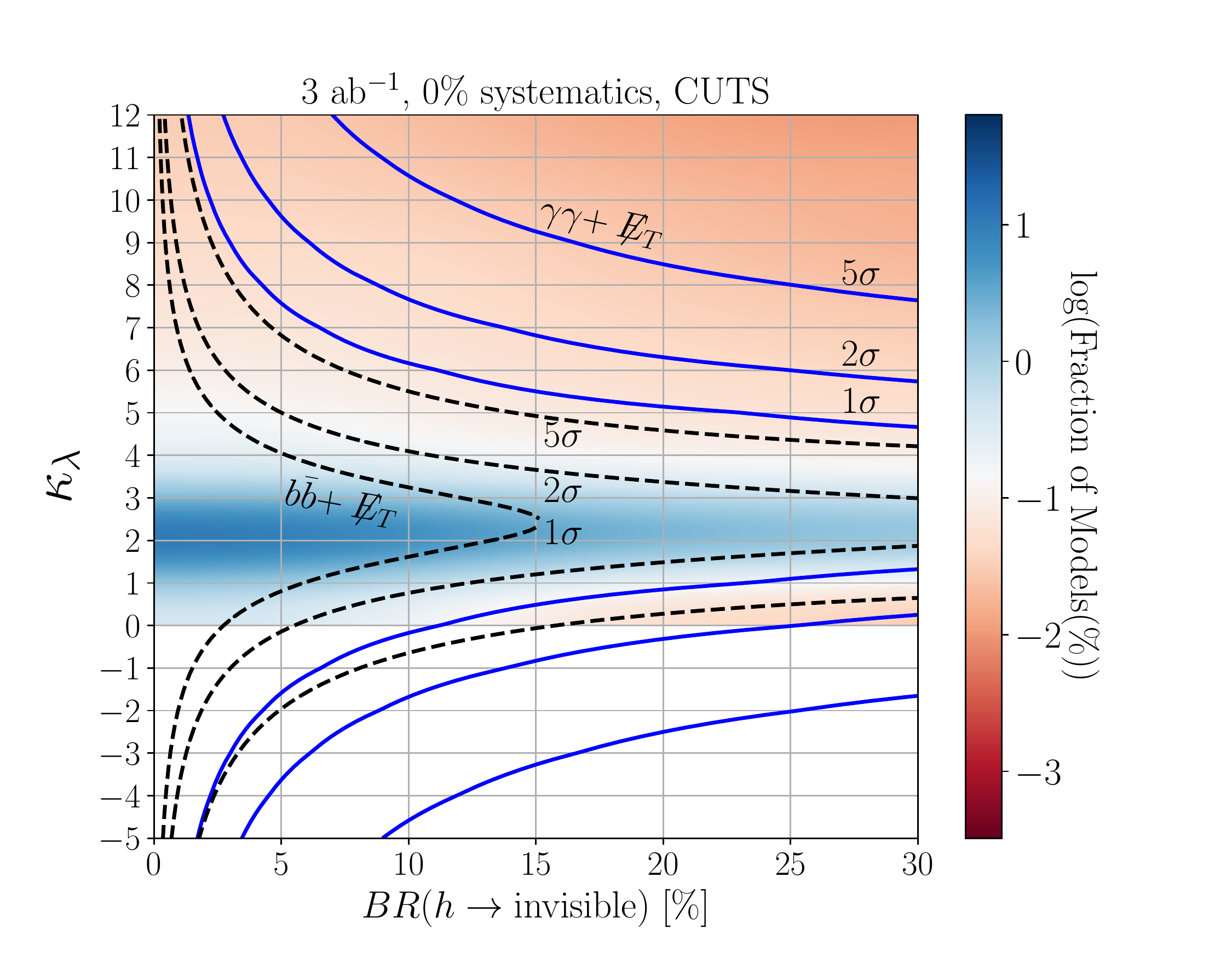}\\ 
  \includegraphics[width=.88\columnwidth]{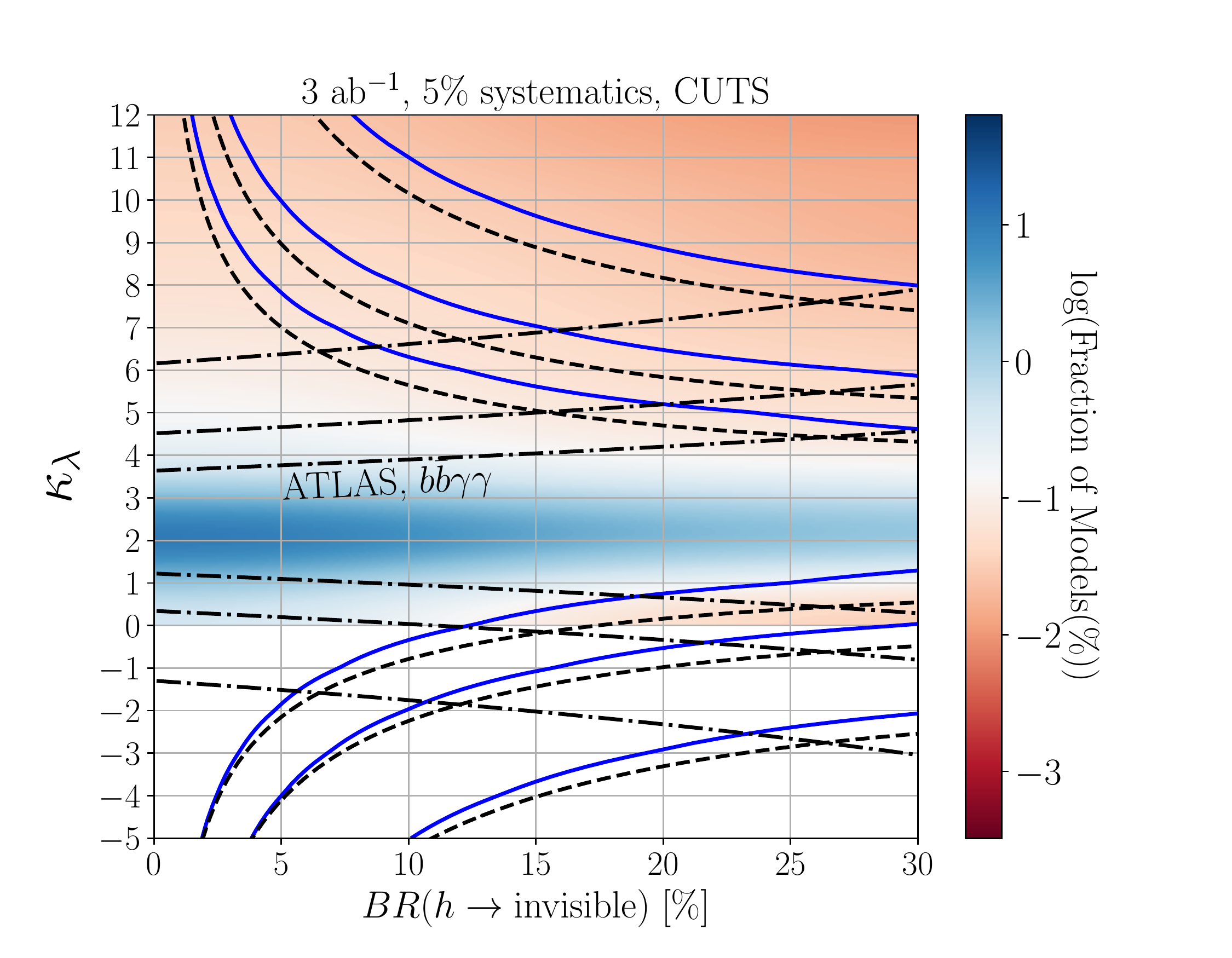}\\
  \includegraphics[width=.88\columnwidth]{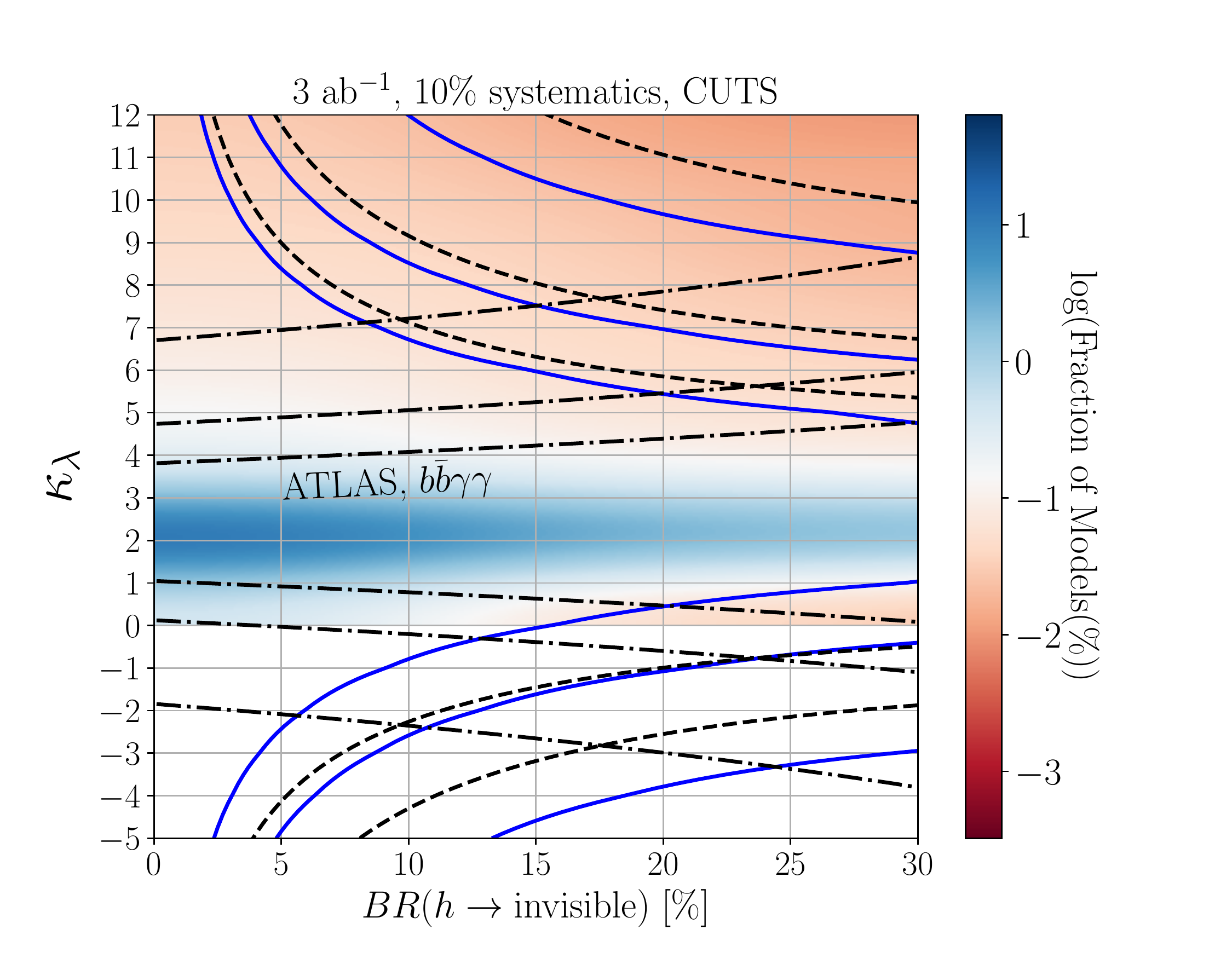}
\caption{\label{fig:res1}
The $1,2$, and $5\sigma$ reaches of the $\aaet$ and $\bbet$ channels for three systematics scenarios in a cut analysis. The cuts of $\aaet$ are optimized for each $\kappa_\lambda$. In the lower panel, we also show the ATLAS~\cite{Aad:2014yja} prospects for $\bbaa$ in the 5 and 10\% systematics cases. The color code denotes the density of xSM models that pass the criteria from Eqs.(\ref{eq:below}--\ref{eq:xmass}).
}
\end{figure}

The production cross section of Higgs pairs depends quadratically on $\lambda_{111}$ where the minimum occurs approximately at $\sim 2.5\lambda_{SM}$, the same point where the signal kinematic distributions get maximally discernible from the backgrounds as we see in Fig.~(\ref{fig:kin}).
Optimizing the cuts raises the cut efficiency and the background rejection for each $\lambda_{111}$ 
and makes the total cross section a more determinant factor across the analysis.

In order to get a straightforward comparison with the results of $\bbet$~\cite{Banerjee:2016nzb}, we fixed the cut efficiency for the SM trilinear coupling and just rescaled the results for other $\lambda_{111}$ by $\mu_{hh}=\sigma(pp\to h_1h_1)_{BSM}/\sigma(pp\to h_1h_1)_{SM}$. This approximation was adopted in Ref.~\cite{Banerjee:2016nzb} to estimate the $5\sigma$ reach and the 95\% CL exclusion limit in the $\mu_{hh}$ {\it versus} $BR(h_1\to\hbox{invisible})$ plane. We borrow the number of signal (for the SM case) and background $\bbet$ final state events after all cuts, which are $S=298$ and $B=11231$ respectively, from Ref.~\cite{Banerjee:2016nzb}. 
For each $\mu_{hh}$ we calculate the corresponding $\kappa_\lambda$.

We show, in Fig.~(\ref{fig:res1}), the $1,2$, and $5\sigma$ reaches after 3 ab$^{-1}$ at the 14 TeV LHC. First, we reproduced the results of Ref.~\cite{Banerjee:2016nzb} using the information available, they are represented by the dashed lines. Note that in the absence of systematic errors, $\bbet$ can exclude the SM trilinear coupling down to $\hbox{BR}(h_1\to\hbox{invisible})\sim 6$\% and all $\kappa_\lambda$ down to $\hbox{BR}_{inv}\sim$ 15\% at 68\% CL. Assuming the current LHC limit of $\hbox{BR}_{inv}=0.19$, trilinear couplings in the interval $|\kappa_\lambda-2.5|\gtrsim 1$ can be excluded at 95\% CL and it is capable to discover di-Higgs production down to $|\kappa_\lambda-2.5|\gtrsim 2$, as we see in the upper panel of fig.~(\ref{fig:res1}). In the statistics dominated scenario, bottoms plus missing energy is more promising than $\aaet$ whose limits and discovery prospects are shown in solid blue lines. We emphasize that these limits take into account just the total number of expected events, in excess of backgrounds, for a given coupling and invisible branching ratio. More stringent limits might be possible in a shape-based analysis of a suitable kinematic distribution just like in those cases where it is possible to reconstruct the di-Higgs mass.


Once we include systematic uncertainties, however, the $\aaet$ channel becomes competitive. Contrary to $\bbet$, whose cuts are fixed, we tuned the cut thresholds of the events of photons plus missing energy aiming the maximization of the signal significance of Eq.~(\ref{eq:sig}) for each integer $\kappa_\lambda$ from $-5$ up to $12$. The optimized cuts for the SM case are quoted in Eq.~(\ref{eq:cut_opt1}) with no systematic errors included
\begin{eqnarray}
& & p_{T_{\gamma_1}} > 33\; \hbox{GeV},\; p_{T_{\gamma_2}} > 20\; \hbox{GeV},\; |M_{\gamma\gamma}-m_h|<5\; \hbox{GeV}\nonumber\\
& & p_{T_{\gamma\gamma}}>20\; \hbox{GeV},\; \met > 27\; \hbox{GeV},\;  \sqrt{\hat{s}}_{min}(0)> 490\; \hbox{GeV}\nonumber \\
& & MT_A > 20\; \hbox{GeV},\; MT_B > 36\; \hbox{GeV},\;\nonumber \\
& & \Delta R_{\gamma\gamma} < 1.5,\; \Delta\phi_{\gamma\gamma} < 1.5,\; |\cos\theta^*| < 0.8,\; N_{jets} > 1\;\; .\nonumber\\
&&
\label{eq:cut_opt1}
\end{eqnarray}

In this case, and for other non-SM $\lambda_{111}$ as well, the cuts found in the optimization process confirm our intuition about the signal rich region in some kinematic variable distributions but, not always. The SM is a good example. The cut of $\Delta R_{\gamma\gamma}<1.5$ seems obvious if we take a look at the bottom panel of Fig.~(\ref{fig:kin}) but not a loose cut on $\met$. The cut in the missing transverse energy turned out to be redundant once a hard cut on $\sqrt{\hat{s}}_{min}(0)$ is imposed as we see in Eq.~(\ref{eq:cut_opt1}) above.

For the SM trilinear coupling, the number of signal events surviving these criteria is 17, and 159 for the total background, corresponding to $1.4\sigma$, and $S/B=0.11$ with no systematics included. The signal-to-background ratio is four times larger than that found in the $\bbet$ study but the signal significance drops by half. As we discussed, this much larger $S/B$ ratio makes $\aaet$ more promising than $\bbet$
when we include systematic effects. In all the subsequent analysis, we tuned the cuts in order to maximize Eq.~(\ref{eq:sig}) with $\varepsilon_{sys}=0$; however, it would be possible to tune the cuts for each systematics level to raise $S/B$ even more.

In the middle panel of Fig.~(\ref{fig:res1}), we display the results for 5\% systematic errors in both signal and background normalizations. The limits of the photons channel move by less than one unit in $\kappa_\lambda$ compared to the 0\% systematic error case, approximately, while the bottoms channel gets much less constraining compared to the statistics dominated scenario. Yet, $\bbet$ is still more constraining by a small margin. Raising the systematic errors to 10\% now flips the situation: the $\aaet$ becomes a better channel to probe di-Higgs production with dark matter decays as we see in the lower panel of Fig.~(\ref{fig:res1}) where we also display the reach of the $\bbaa$ channel using the results of the ATLAS Collaboration~\cite{Aad:2014yja} projecting the prospects of the 14 TeV LHC and 3 ab$^{-1}$ in a cut-and-count study. To calculate the corresponding $\kappa_\lambda$ of a given invisible branching ratio, we multiplied the number of SM signal events quoted by ATLAS by $(1-\hbox{BR}_{inv})^2$ and then obtained the $\kappa_\lambda$ that would enhance the significance to $1,2$, and $5\sigma$. 
As a final remark, note that $\bbaa$ probes more significant regions in $\kappa_\lambda$ than the other two channels for all the relevant invisible branching ratios in the cut analysis.



%
\begin{figure}[b]
  \includegraphics[width=.9\columnwidth]{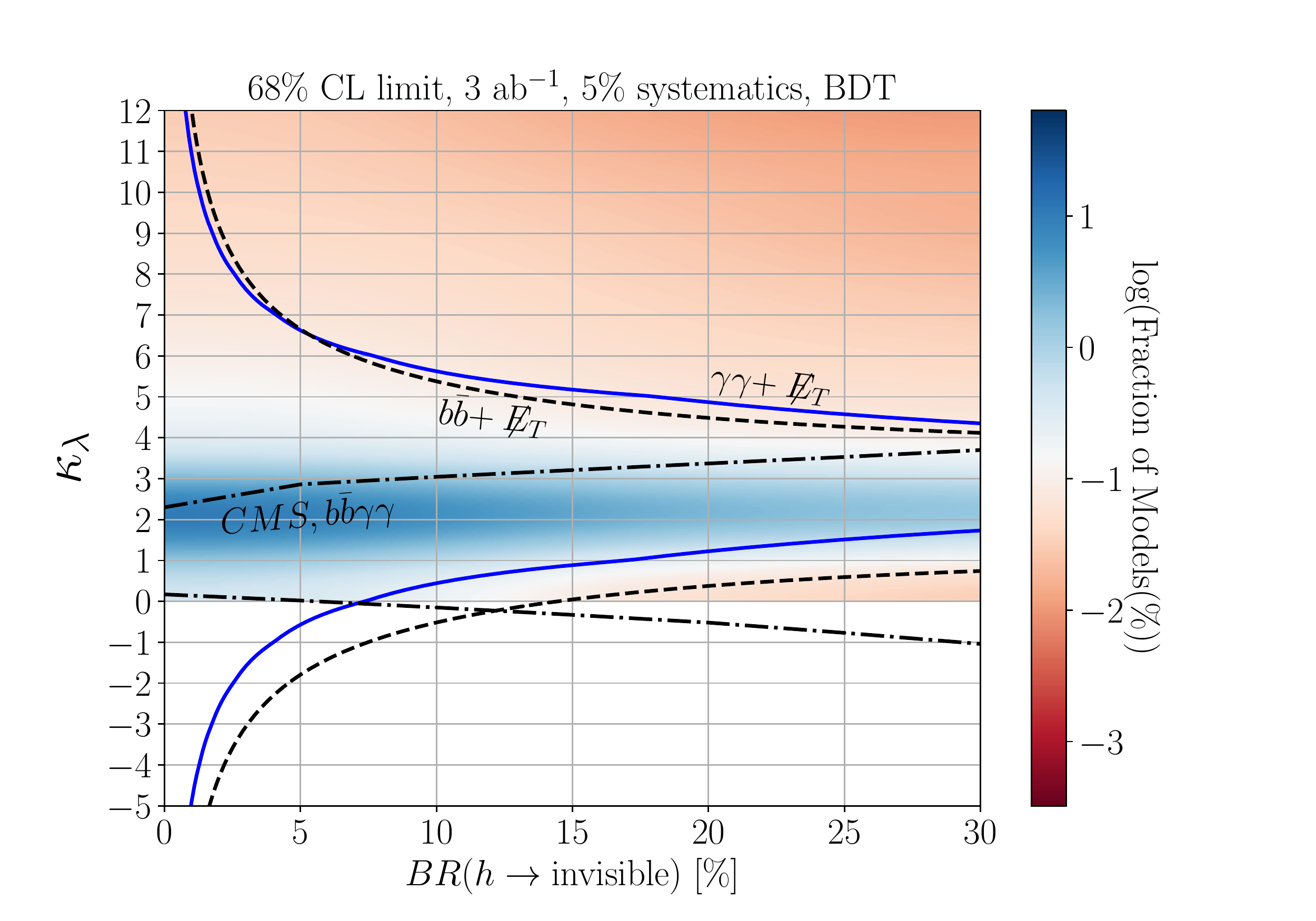}\\ 
  \includegraphics[width=.9\columnwidth]{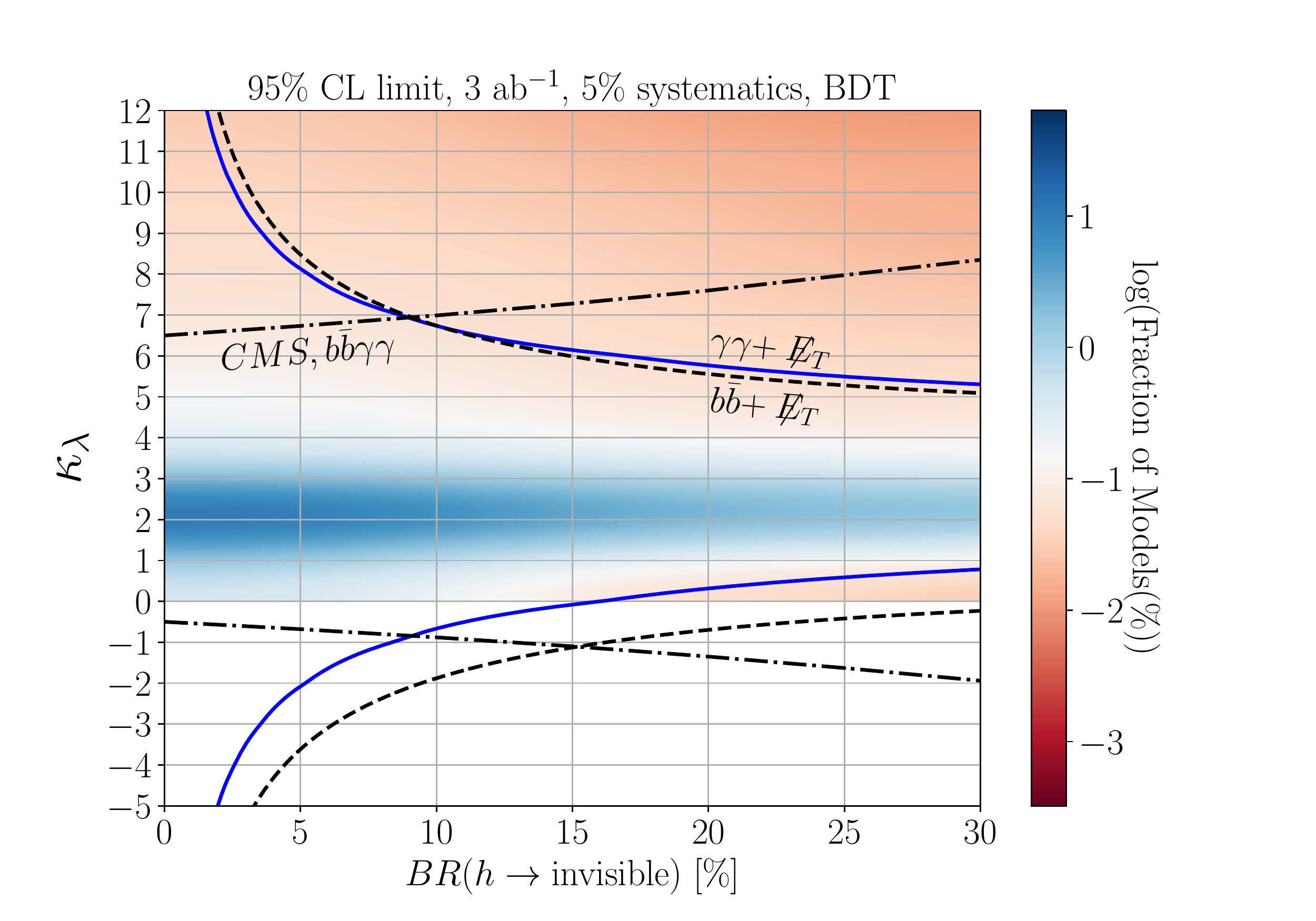}\\
  \includegraphics[width=.9\columnwidth]{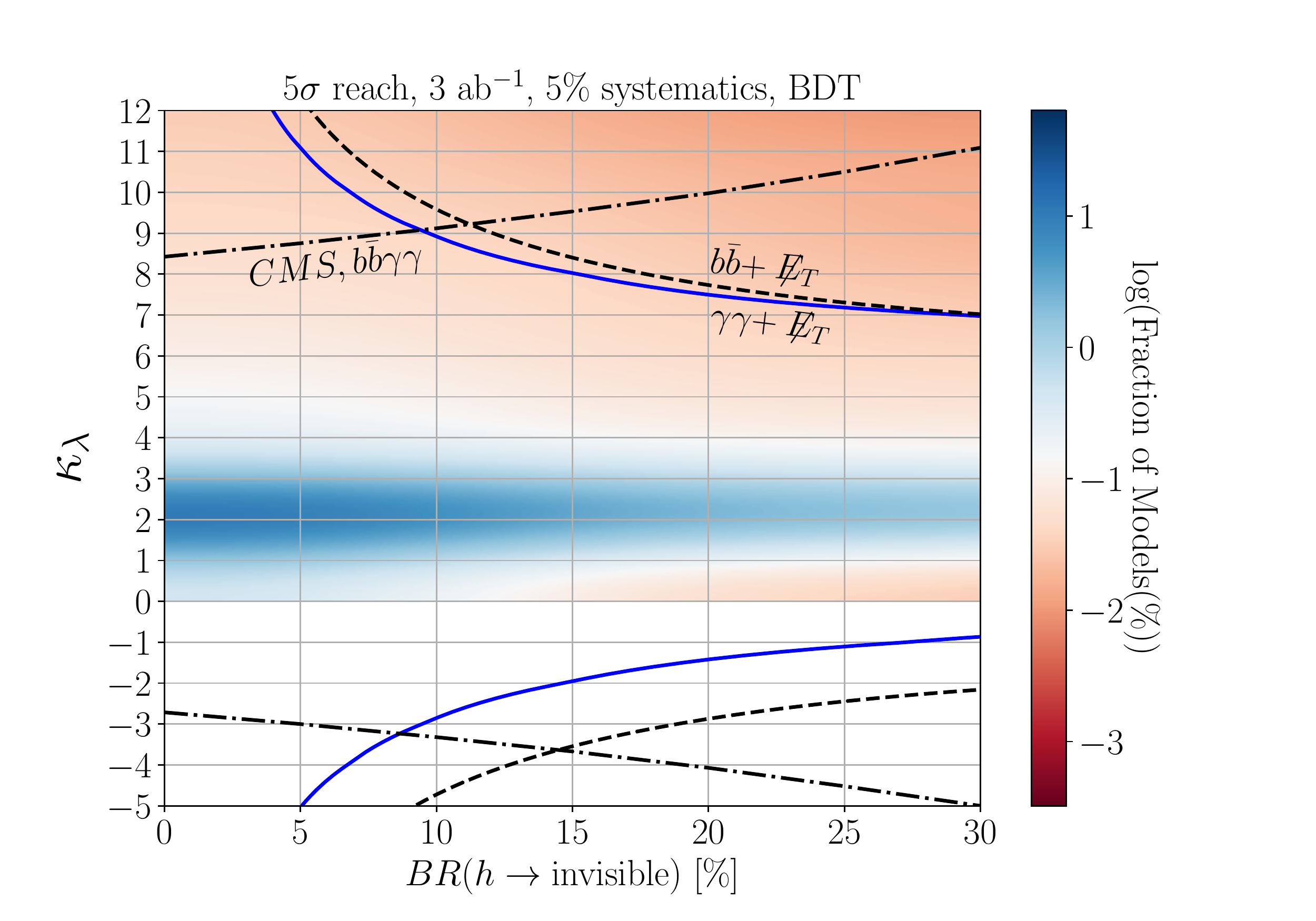}
\caption{\label{fig:res2}
The $68,95$\% CL limits, and $5\sigma$ reaches of the $\aaet$ and $\bbet$ channels for 5\% systematics scenarios in the BDT analysis. The cuts of the dark and bright channel are optimized for each $\lambda$. In all panels, we also show the CMS~\cite{Cepeda:2019klc} prospects in the $\bbaa$.
}
\end{figure}

\section{\label{sec:BDT} BDT Analysis} \label{dihiggs}

Although it improves the reach of the LHC for double Higgs production with one invisibly decaying Higgs boson, the $\aaet$ channel is still not competitive with $\bbaa$ in constraining the Higgs trilinear coupling in the presence of an invisible decay mode as we discussed in the last section. The prospects change when we perform a multivariate analysis.


We used BDTs, as implemented in \texttt{XGBoost}~\cite{xgb}, to better classify our signal and background events. Just like the cut-and-count analysis, we tuned the cuts but also the BDT hyperparameters in a  joint optimization of the signal significance of Eq.~(\ref{eq:sig}) with no systematics which were included just in the final computation of the statistical significance. The optimization of a dedicated classifier was performed for all the trilinear couplings corresponding to an integer $\kappa_\lambda$ from $-5$ to $12$ taking into account the changes in the fifteen kinematic distributions described in Section~\ref{sec:cuts}.

We split our dataset, of around 100 thousand events for each one of the three background classes and the signal class (totaling 400k events), in two equal parts, one for testing and the other for training. We used \texttt{CutOptimize} to perform 200 iterations in the joint parameters space to maximize the significance. In each iteration, we randomly split the data 5 times in training and test sets of equal parts and calculated the BDT outputs of each class, estimating their distributions.
These distributions were then used to place a final cut in order to better separate the signal and the backgrounds. The median of the five significance signals gives a final significance which constituted the optimization objective of the maximization algorithm. Once the best cuts and BDT hyperparameters were found, we perform a final 10-fold cross validation by randomly splitting the data set in train/test sets of equal size to estimate the final significance and its variance. In all the final results, the variance of the signal significance was small indicating the robustness of the best parameters found.

 A feature importance analysis using the \texttt{SHAP}~\cite{NIPS2017_7062, lundberg2018consistent, shap} package, shows that $\met$, $M_{\gamma\gamma}$, $\sqrt{\hat{s}}_{min}(0)$, $MT_B$ and $N_{jets}$ are the most important variables for the BDT classification for the majority of trilinear couplings.


We took the number of signal and backgrounds events for $\bbet$ from Ref.~\cite{Banerjee:2016nzb} after their BDT analysis using \texttt{TMVA}, namely, $S=593$ and $B=19466$ but again fixing the cut efficiency for other $\kappa_\lambda$ values and just scaling the signal significance by the total cross sections of the signal. It must be emphasized that this approach is only an approximation since the cut efficiency of the signal varies with $\kappa_\lambda$.

We also took the prospects for the $\bbaa$ channel from a recent CMS study~\cite{Cepeda:2019klc} of the di-Higgs production at the HL-LHC using BDTs to purify the samples and using a parametric maximum likelihood fit to photons, $m_{\gamma\gamma}$, and bottoms, $m_{b\bar{b}}$, masses distributions. In order to estimate the couplings that can be probed with reduced branching ratios into $b\bar{b}$ and $\gamma\gamma$ due to the presence of the dark matter decay, we just rescaled the signals significance of the CMS work by $(1-BR_{inv})^2$ and looked for the $\kappa_\lambda$ with that new, smaller, significances.

In the upper panel of Fig.~(\ref{fig:res2}), we display the 68\% CL limits of the three search channels in the $\kappa_\lambda\times \hbox{BR}_{inv}$ plane for 5\% systematics in both signal and background normalizations.
First of all, we note that the optimization process of $\aaet$ is able to find parameters for which the results in terms of signal significance vary smoothly as a function of $\kappa_\lambda$ and $\hbox{BR}_{inv}$. This behavior indicates that the optimization algorithm finds the path of the points of the maximum of the objective function in the multidimensional parameters space. The $\aaet$ and $\bbet$ channels perform nearly the same but, again, the bottoms channel is not optimized. The prospects using $\bbaa$ events are the better at this confidence level for all the relevant invisible Higgs branching ratios but, as expected, these limits soften as $\hbox{BR}_{inv}$ gets larger. The color code of the heatmaps superimposed on the plots of Fig.~(\ref{fig:res2}) shows the density of models respecting all constraints of Eqs.~(\ref{eq:below}--\ref{eq:xmass}) in the $\kappa_\lambda$ {\it versus} $\hbox{BR}_{inv}$ plane. The blue region concentrates the majority of models. Part of that region can be probed in $\bbaa$ and $\aaet$ at 68\% CL, as we see in the upper panel. The blank regions do not contain viable points of xSM, but we show them once they might constrain other interesting models with negative shifts in the trilinear coupling.

The middle panel of Fig.~(\ref{fig:res2}) presents the 95\% CL limits. The dark matter channels now probe a region moved approximately one unit in $|\kappa_\lambda|$ upwards compared to the 68\% CL case. However, the $\bbaa$ limits, as obtained by the CMS Collaboration, soften more intensely. Interestingly, a complementarity now arises where $\bbaa$ probes small invisibly Higgs decays up to $\sim 8$\%, while the other two channels constrain higher branching ratios. Only trilinear couplings either larger than $5.5\lambda_{SM}$ or smaller than $0.5\lambda_{SM}$ can be observed at this confidence level.

In the last panel of Fig.~(\ref{fig:res2}), we show the discovery prospects assuming, again, a 5\% systematics. Now, $\aaet$ performs better than $\bbet$ for all invisible branching ratios. This channel is also better than $\bbaa$ for $\hbox{BR}_{inv}\gtrsim 8$\%. We estimate that it is possible to discover di-Higgs production and decay to bright and dark states for $\lambda_{111}\gtrsim 7.7\lambda_{SM}$ in xSM, and $\lambda_{111}\lesssim -1.4\lambda_{SM}$ in models with negative trilinear shifts, for $\hbox{BR}_{inv}<19$\%.

\begin{figure}[t]
  \includegraphics[width=1.05\columnwidth]{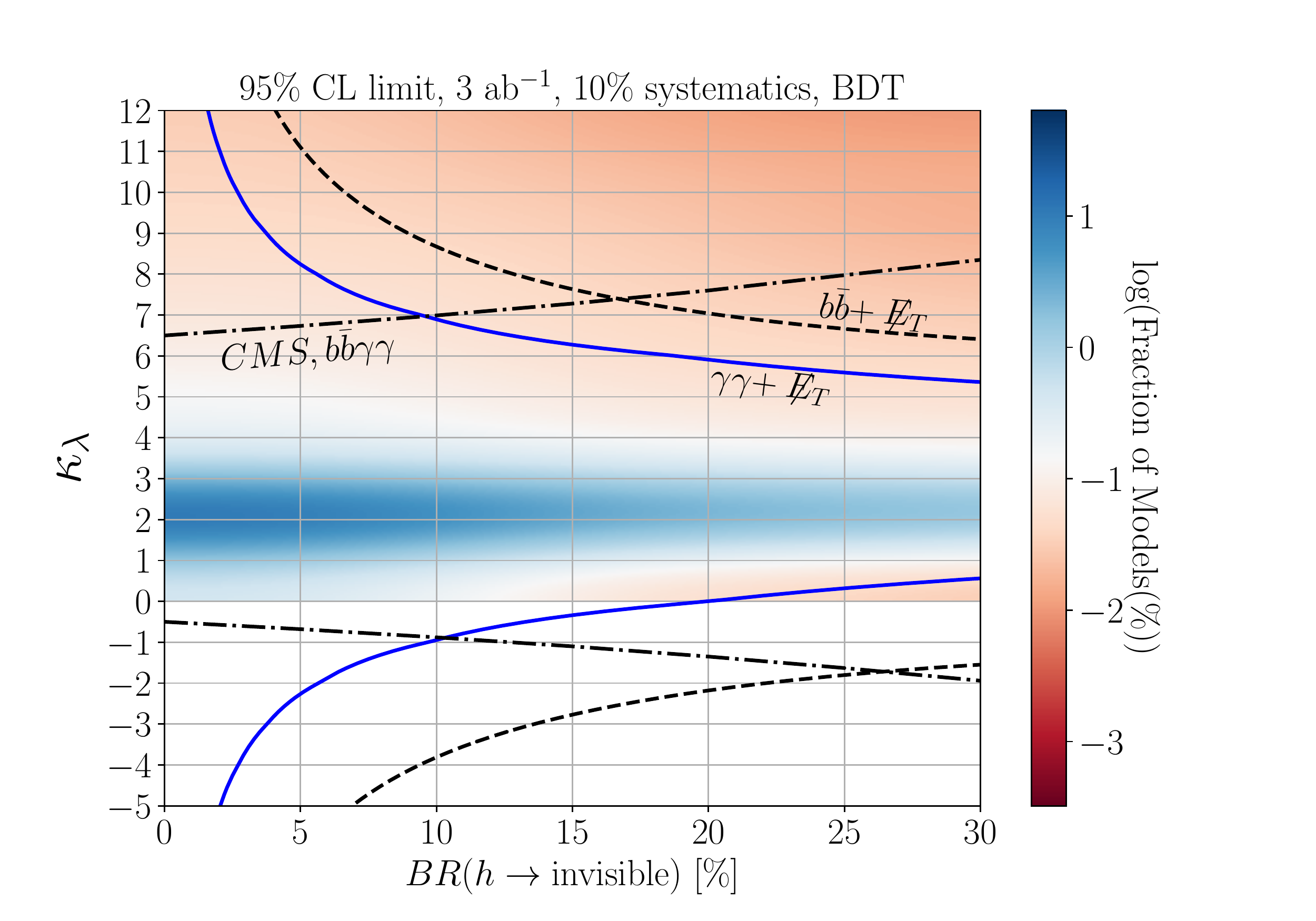}
\caption{\label{fig:res3}
The $95$\% CL limits of the $\aaet$ and $\bbet$ channels for 10\% systematics scenarios in the BDT analysis. The cuts of the dark and bright channel are optimized for each $\kappa_\lambda$. We also show the CMS~\cite{Cepeda:2019klc} prospects in the $\bbaa$ channel.
}
\end{figure}

Finally, we raise the systematics to 10\% and estimate the prospects of the HL-LHC at the 95\% CL using BDTs. In Fig.~(\ref{fig:res3}), we see that $\kappa_\lambda$ down to 6 can be probed for $\hbox{BR}_{inv}=19$\% with $\aaet$ events and a complementarity with $\bbaa$ occurs around $\hbox{BR}_{inv}=10$\%. Compared to these topologies, raising the systematics level has a more deleterious effect on the reach of $\bbet$ channel.

%
%

\section{\label{sec:conclusions} Conclusions}

The production of two Higgs bosons is of prime importance to understand the scalar sector of the Standard Model. Determining
the self-interactions of the Higgs boson might reveal a more profound structure of the scalar potential with clues to important
open questions as the stability of the vacuum and an electroweak phase transition that has driven the baryon matter asymmetry,
for example.

Along with the nature of the Higgs sector, one of the most obvious targets of colliders is dark matter. If just like the other known particles, the
dark matter has its mass generated through the Higgs mechanism, it is plausible to observe a Higgs-DM interaction. Especially, if the Higgs boson decays
to DM, new possibilities to search for double Higgs production may open up. On the other hand, the number of events from decay channels involving only the SM particles
are expected to be reduced by a factor of $(1-\hbox{BR}_{inv})^2$, which is around 2/3 given the current LHC bound on the invisible decay width of the Higgs. All this makes the study
of double Higgs production and decay to DM an interesting and timely task.

In this work, we investigated the di-Higgs production in three complementary channels: $\aaet$, $\bbet$ and $\bbaa$. The prospects of the LHC in the dark
and bright channel $\aaet$ were estimated for the first time in both a cut-and-count and multivariate analysis using a dedicated algorithm to optimize their
signal significances. The estimates for $\bbet$ and $\bbaa$ were taken from the literature adapting to the inclusion of the Higgs decays to DM when necessary.

Our approach aimed at the singlet scalar extension of the SM, the xSM, which is the simplest extension of the SM that leads to first order EWPT. Under the assumptions that we made, namely, a massive new Higgs boson with small mixing with the SM Higgs boson, a more or less model-independent estimate follows and an effective field theory approach is reliable. In order to keep new ingredients to the SM to a minimum, we
 augmented xSM with a fermionic DM coupling to the new scalar. Concerning the DM sector, we demanded points are respecting both the latest XENON1T bounds. Moreover, for DM masses not too close to Higgs mediator thresholds the relics abundance bound can also be evaded.

For the xSM, we found that portions of the parameters space that will give us significant deviation of the trilinear Higgs coupling relative to the SM value can be probed by, at least, one of the decay channels studied here. 
The $\aaet$ channel, in particular, becomes a better option than $\bbet$ once systematic uncertainties are taken into account. Moreover, the dark and bright channel presents better prospects for exclusion and discovery than $\bbaa$ for $\hbox{BR}_{inv}$ as low as $\sim 10$\% in the multivariate analysis using BDTs. Overall, all three channels complement each other in the region $0<\hbox{BR}_{inv}<19$\% and $-5\leq\kappa_\lambda\leq 12$, which is still allowed by the data. Our results also show that those parameters of xSM with a trilinear coupling close to the SM and an invisible Higgs branching ratio smaller than $\sim 15$\% will be tough to be probed at the LHC relying only on shifts of the SM Higgs self-interactions. Scenarios with not so heavy new Higgs bosons are potentially more easily accessed in colliders and a complementary study across many search channels like this one is due.

\section*{Acknowledgments}
AA thanks Conselho Nacional de Desenvolvimento Cient\'{\i}fico (CNPq) for
its financial support, grant 307265/2017-0. FSQ acknowledges support from CNPq
grants 303817/2018-6 and 421952/2018-0, UFRN, MEC
and ICTP-SAIFR FAPESP grant 2016/01343-7. TG is supported by U. S. Department of Energy grant de-sc0010504.

\bibliography{mybib}
\end{document}